\documentclass[useAMS,usenatbib]{mn2e}

\def\rfr#1{eq.(\ref{#1})}

\def\eqi{\begin{equation}}
\def\eqf{\end{equation}}
\def\eqia{\begin{eqnarray}}
\def\eqfa{\end{eqnarray}}

\def\rp#1#2{{#1\over#2}}

\def\lb#1{\label{#1}}

\def\ga{\gamma}

\def\bm#1{{\mbox{\boldmath$#1$\unboldmath}}}

\def\to{$T_{\rm obs}$}

\title[post-Newtonian correction to the orbital period and its possible measurement]{On the possibility of measuring the
post-Newtonian gravitoelectric correction to the orbital period of
a test body in a Solar System scenario}
\author[L. Iorio]{L. Iorio$^{1}$\thanks{E-mail:
lorenzo.iorio@libero.it (AVR)} \\
$^{1}$Dipartimento di Fisica dell'Universit$\grave{\rm a}$ di
Bari, Via Amendola 173, 70126, Bari, Italy}
\begin{document}


\pagerange{\pageref{firstpage}--\pageref{lastpage}} \pubyear{2005}

\maketitle

\label{firstpage}

\begin{abstract}
The possibility of measuring the post-Newtonian gravitoelectric
correction to the orbital period of a test particle freely
orbiting a spherically symmetric mass in the Solar System is
analyzed. It should be possible, in principle, to detect it for
Mercury at a  precision level of 10$^{-4}$. This level is mainly
set by the unavoidable systematic errors due to the mismodelling
in the Keplerian period which could not be reduced by accumulating
a large number of orbital revolutions. Future missions like
Messenger and BepiColombo should allow to improve it by increasing
our knowledge of the Mercury's orbital parameters. The
observational  accuracy is estimated to be $10^{-4}$ from the
knowledge of the International Celestial Reference Frame (ICRF)
axes. It could be improved by observing as many planetary transits
as possible.
It is not possible to measure such an effect in the gravitational
field of the Earth by analyzing the motion of artificial
satellites or the Moon because of the unavoidable systematic
errors related to the uncertainties in the Keplerian periods. In
the case of some recently discovered exoplanets  the problems come
from the observational errors which are larger than the
relativistic effect.
\end{abstract}

\section{The post-Newtonian gravitoelectric correction to the orbital period}
The geodesic motion of a test particle in the gravitational field
of a spherically symmetric body of mass $M$ is considered. In the
framework of the Einstein's General Theory of Relativity the
post-Newtonian gravitoelectric acceleration of order
$\mathcal{O}(c^{-2})$ experienced by the test particle is
\citep{cazonga} \eqi \bm a^{(\rm GE)}=\rp{GM}{c^2
r^3}\left\{\left[4\rp{GM}{r}-(\bm v\cdot\bm v) \right]\bm r+4(\bm
r\cdot\bm v)\bm v\right\},\lb{age}\eqf where $G$ is the Newtonian
gravitational constant, $c$ is the speed of light in vacuum, $\bm
r$ and $\bm v$ are the position and velocity vectors,
respectively, of the test particle,
$r$ is the standard isotropic radial coordinate \footnote{It is
the coordinate used in the force models of the post-Newtonian
equations of motions \citep{est71} adopted for the computations of
the planetary ephemerides by, e.g., the Jet Propulsion Laboratory
(JPL). }, not to be confused with the Schwarzschild radial
coordinate $r^{'}=r[1+GM/(2c^2 r)]^2$. The gravitoelectric
acceleration of \rfr{age} is the same which is responsible for the
Einstein secular advance of the perihelia \citep{ein15} of the
Solar System planets which are known at a $10^{-4}$ level of
accuracy \citep{pit01a, pit01b}.

We are interested in the consequences of \rfr{age} on the orbital
period of the test particle in order to see if they can be
detected in suitably designed experiments in the Solar System
arena. Let us first consider circular orbits for which $\bm
r\cdot\bm v=0$ and $r=a$, where $a$ is the semimajor axis of the
orbit; indeed, the orbits of the planets and of many of the best
tracked Earth artificial satellites have rather small
eccentricities $e$. Subsequent sensitivity analyses will show the
extent to which this approximation can be considered adequate. In
this case the acceleration given by \rfr{age} is directed radially
and the radial equation of motion reduces to \eqi
\rp{v^2}{a}=\frac{GM}{a^2}-\rp{GM}{c^2 a^3}\left[4\rp{GM}{a}-v^2
\right].\eqf With $v=a d\phi/dt$, the time required for the
azimuthal angle $\phi$ to pass from 0 to 2$\pi$ becomes \eqi
P\equiv P^{(0)}+P^{(\rm
GE)}=2\pi\sqrt{\rp{a^3}{GM}}+\rp{3\pi}{c^2}\sqrt{GMa},\lb{perge}\eqf
where $P^{(0)}$ is the unperturbed Keplerian period
Note that \rfr{perge}
is consistent with
%
the result of \citep{mas01}. Contrary to the gravitomagnetic
correction to the Keplerian orbital period \citep{mas01}, which is
generated by the proper angular momentum $\bm J$ of the central
mass, $P^{(\rm GE)}$ is insensitive to the direction of motion of
the test particle around its orbit and to its inclination with
respect to the equatorial plane of the central body.
\section{A tentative error budget}
Is it possible to measure $P^{(\rm GE)}$ by analyzing the motion
of the planets around the Sun? Although their eccentricities are
rather small being of the order of $10^{-2}-10^{-3}$, apart from
Mercury for which $e=0.2056$, it turns out that, as we will show
below, the available observational accuracy does not allow to
neglect the impact of the non-circularity of their orbits. For
$e\neq 0$ the exact expression of the post-Newtonian
gravitoelectric correction is \citep{mas01} \eqi P^{(\rm
GE)}=\rp{3\pi}{c^2}\sqrt{GMa}\left[3-2\rp{\sqrt{1-e^2}}{(1+e\cos
f_0 )^2}\right]\lb{tutta}\eqf where
$f_0$ is the true anomaly at the initial epoch of the planet.
%
%
%
%
The factor \eqi 3-\rp{2\sqrt{1-e^2}}{(1+e\cos
f_0)^2}\lb{fattore}\eqf runs from 1.65 for $f_0=0^{\circ}$ to
-0.10 for $f_0=180^{\circ}$; in the case of Venus, which has
$e=0.00677$, it changes from 1.02 to 0.97. In the following, in
order to sketch a sensitivity analysis, we will work with the
initial epoch for which the factor of \rfr{fattore} is equal to 1,
i.e. we will assume \rfr{perge}.
\subsection{The observational sensitivity}\lb{sensi}
The azimuthal angle $\phi$ is the usual right ascension $\alpha$
of a spherical coordinate system in an astronomical reference
frame.

We will refer to the International Celestial Reference Frame
\footnote{It is the realization of the International Celestial
Reference System (ICRS) \citep{ari95} by means of the estimates of
a set of extragalactic sources \citep{ma98, fey04}.} (ICRF)
\citep{mcc04}. Moreover, the right ascension is one of the direct
observables in planetary motions. Since we are interested in the
times when the right ascension of a planet crosses the $\{x,z\}$
plane of ICRF, i.e. when it is zero, it is of the utmost
importance to see if the present uncertainty in the stability of
the ICRF's axes would allow for a detection of the gravitoelectric
correction or if it is too large and would mask it. We can
reasonably assume that the right ascension of a planet advances
under the effect of the post-Newtonian gravitoelectric force over
an angular interval of $\overline{\alpha}=2\pi$, by a quantity
$\delta\alpha$ defined by
\eqi\rp{\delta\alpha}{\overline{\alpha}}\sim \rp{P^{(\rm
GE)}}{P^{(0)}}=\rp{3GM_{\odot}}{2c^2 a}\lb{daa},\eqf where $a$ is
the semimajor axis. From \rfr{daa} it is apparent that the inner
planets have to be considered in order to maximize the effect. For
Mercury, whose semimajor axis amounts to $a=0.38709893$ A.U.,
$P^{(\rm GE)}=0.29$ s (if the numerical factor of \rfr{fattore} is
assumed equal to 1) and $P^{(0)}=7.60055184\times 10^6$ s;
\rfr{daa} yields $\delta\alpha/\overline{\alpha}=3.8\times
10^{-8}$. Then, $\delta\alpha= 50$ milliarcseconds (mas in the
following). According to \citep{mcc04}, the uncertainty in the
ICRF axes amounts to $\sigma_{\alpha}=0.02$ mas. This yields an
accuracy of $4\times 10^{-4}$.
%
%
Then it appears clear that with such an accuracy\footnote{It maybe
interesting also to note that such an accuracy would not allow for
a detection of the gravitomagnetic correction to the Keplerian
period $2\pi J_{\odot}/c^2 M$ which, in the case of the Sun,
amounts to $4\times 10^{-6}$ s by assuming \citep{pij03}
$J_{\odot}=1.9\times 10^{41}$ kg m$^2$ s$^{-1}$.}
the circular case approximation cannot be considered adequate.

Such evaluations hold for a single orbital revolution only; a
measurement like that proposed here must be performed over a large
number of planetary transits. This would greatly increase the
observational accuracy.

Our knowledge of the orbital motion of Mercury will improve thanks
to the future missions Messenger \footnote{See on the WEB
http://messenger.jhuapl.edu/ and
http://discovery.nasa.gov/messenger.html}, which has been launched
in the summer 2004 and whose encounter with Mercury is scheduled
for 2011, and, especially \footnote{While the spacecraft
trajectory will be determined from the range-rate data, the
planet's orbit will be retrieved from the range data
\citep{mil02}. In particular, the determination of the planetary
centre of mass is important to this goal which can be better
reached by a not too elliptical spacecraft's orbit. The relatively
moderate ellipticity of the planned 400$\times$ 1500 km polar
orbit of BepiColombo main orbiter, opposite to the much more
elliptical path of Messenger, is, then, well adequate.},
BepiColombo \footnote{See on the WEB
http://sci.esa.int/science-e/www/area/index.cfm?fareaid=30}, which
is scheduled to fly in 2010-2012. A complete error analysis for
the range and range-rate measurements to BepiColombo can be found
in \citep{ies01}. According to them, a two orders of magnitude
improvement in the Earth-Mercury range, which is accurate to
hundreds of meters now, should be possible. According to a more
conservative evaluation by E.M Standish (JPL)(Standish, private
communication 2004), improvements in the Mercury's orbital
parameters might amount to one order of magnitude.
In regard to the proposed test, the Mercury transits which could
benefit from Messenger and BepiColombo will be of the order of 4,
since both the spacecraft should nominally orbit Mercury for 1
year and the orbital period of Mercury around the Sun is almost
equal to 88 days.

\subsection{Some systematic errors}
\subsubsection{The errors in the Keplerian period}
The value of the Keplerian period, evaluated from the estimated
semimajor axis $a$ and Sun's $GM$, must be subtracted from the
data record in order to single out the post--Newtonian effect (and
the other classical and post-Newtonian perturbations, of course,
which, in this case, would represent the noise). Then, let us see
if the systematic errors in the Keplerian period are smaller than
the post-Newtonian gravitoelectric correction which we are
interested in. This is a very important point because, opposite to
the observational error, the impact of this source of systematic
bias could not be reduced by observing many orbital revolutions.
We have \eqi \sigma_{P^{(0)}}\leq
3\pi\sqrt{\rp{a}{GM_{\odot}}}\sigma_{a}+\pi\left(\rp{a}{GM_{\odot}}\right)^{3/2}\sigma_{GM_{\odot}}.
\eqf By assuming\footnote{In fact, the formal, statistical error
is 0.187 m: in this case $\sigma_{P^{(0)}}\leq 3\times 10^{-5}$ s.
However, the real error bound in $a$ might be one order of
magnitude larger. Note also that according to some authors the
solar $GM$ should be considered as fixed.} $\sigma_{{a}^{\rm
Merc}}=1.87$ m \citep{pit01a} and $\sigma_{GM_{\odot}}=8\times
10^9$ m$^3$ s$^{-2}$ \citep{sta95}, we have $\sigma_{P^{(0)}}\leq
3\times 10^{-4}$ s + $2\times 10^{-4}$ s. The NASA Messenger and
the future ESA BepiColombo missions should allow to further reduce
$\sigma_{P^{(0)}}$ because it will yield a better knowledge of the
parameters of the Sun-Mercury system.

The effect of the quadrupole mass moment $J_{2\odot}$ of the Sun
would induce a correction \eqi P^{(J_{2\odot})}\sim-\rp{6\pi
R_{\odot}^2 J_{2\odot}}{\sqrt{GM_{\odot}a}}.\eqf By assuming the
range $J_{2\odot}=(2\pm 0.4)\times 10^{-7}$ \citep{pir03}, it
amounts to $(6\pm 1.2)\times 10^{-4}$ s. Note this error should be
reduced to $\sim 10^{-5}$ s if $J_{2\odot}$ will be measured with
an accuracy of the order of $10^{-9}$ by BepiColombo.
These sources of constant bias, which are at the same level of the
one-revolution observational error previously outlined, set the
limit of the obtainable accuracy over a record spanning over many
planetary revolutions.
\subsubsection{The direct planetary secular perturbations}
Another source of perturbations on the Mercury's right ascension
is represented by the gravitational perturbations induced by the
other major bodies of the Solar System. Let us calculate the
secular effects induced by some of the other planets. The
perturbative effect of the planet $m_j$ on the planet $m_i$ is
given by \citep{boc99} \eqi\mathcal{R}^{(\rm
planets)}=Gm^j\left(\rp{1}{|\bm r^i -\bm r^j |}-\rp{\bm r^i\cdot
\bm r^j}{r^j{^3}}\right).\lb{bocpuc}\eqf It turns out that the
second term in \rfr{bocpuc} does not induces secular
perturbations. After expressing the first term of \rfr{bocpuc} in
terms of the orbital elements of the i-th and j-th bodies and
averaging it over one period of the mean longitudes $\lambda^i$
and $\lambda^j$ it can be obtained, for the largest contribution
which does not contain the terms of second order in the
eccentricities and the inclinations \eqi P^{(\rm
planets)}=-\rp{4\pi Gm^j}{n^{i^3} a^{j^3}}.\lb{plan}\eqf The
nominal values of \rfr{plan} are larger than the gravitoelectric
effect: the major contributions come from Jupiter (-5.97 s), Venus
(-5.70 s), the Earth (-2.64 s) and Saturn (-0.29 s). If the
estimates of \rfr{plan} are correct, this would not pose problems
because the systematic errors induced by $\sigma_{{a^j}}$ and
$\sigma_{Gm^{j}}$ would be far smaller than the gravitoelectric
effect. Indeed, by assuming for Jupiter $\sigma_{{a^{\rm
Jup}}}=675.1$ m \citep{pit01a} and \footnote{See on the WEB
http://ssd.jpl.nasa.gov/sat$\_$gravity.html} $\sigma_{Gm^{\rm
Jup}}=2\times 10^9$ m$^3$ s$^{-2}$ \citep{jac03} the uncertainty
in its contribution would amount to $10^{-7}$ s only.
The nominal values of the corrections induced by Jupiter allow to
guess that the impact of many of the asteroids, whose masses are
not accurately known \citep{sta02}, should be negligible: indeed,
from \rfr{plan} we can assume that $P^{\rm (asteroid)}/P^{(\rm
Jup)}\sim m^{(\rm asteroid)}/m^{(\rm Jup)}$.
\subsubsection{The indirect errors in the Keplerian period due to the short-term variations of the semimajor axis}
A possible source of aliasing for Mercury could be represented by
the indirect perturbations \eqi\Delta
P^{(0)}=3\pi\sqrt{\frac{a}{GM_{\odot}}}\Delta a=2\times 10^{-4}\
{\rm s\ m^{-1}}\times \Delta a\eqf induced by short--periodic
effects on $a$, i.e. those effects which are not averaged over one
orbital revolution of the various planets. They are
\begin{itemize}
\item The indirect effects on $P^{(0)}$ induced by the high--frequency
perturbations on Mercury's semimajor axis due to post--Newtonian
gravity itself. They can be calculated from \rfr{age} and the
Gauss perturbative equation for the rate of the semimajor axis. It
turns out that they are $$\Delta a^{(\rm
GE)}=\frac{GM_{\odot}e}{c^2(1-e^2)^2} [14(\cos f_0-\cos
f)+$$\eqi+10e(\cos^2 f-\cos^2 f_0 )] +{\mathcal{O}}(e^3).\eqf For
Mercury their nominal amplitudes are of the order of 4 km; by
assuming $\sigma_{GM_{\odot}}=8\times 10^9$ m$^3$ s$^{-2}$ the
error in them is of the order of $10^{-7}$ m. We can, then,
conclude that the indirect effects due to the post-Newtonian
short-period shifts in the semimajor axis are negligible.

\item The indirect effects on $P^{(0)}$ induced by the high--frequency
perturbations on Mercury's semimajor axis due to the classical
N--body perturbations.
In order to get just some orders of magnitude, let us see what
could be the perturbation induced on the semimajor axis of Mercury
by a planet whose orbital elements will be marked with $^{'}$. It
turns out that, for the term of lowest degree
$|l|+|l^{'}|+|m|+|m^{'}|$ which must be $\leq 2$ because in the
Lagrange planetary equations only the first partial derivatives
appear, we can write
$$d a\propto \frac{2}{n^2 a}
\frac{GM^{'}}{a^{'}}\sum_{jj^{'}ll^{'}mm^{'}}je^{|l|}e^{'
|l^{'}|}\sin^{|m|}
i\sin^{|m^{'}|}i^{'}\times$$\eqi\times\sin(j\lambda+j^{'}\lambda^{'}+l\varpi+l^{'}\varpi^{'}+m\Omega+m^{'}\Omega^{'})dt
,\lb{short}\eqf where $j\neq 0$, $\varpi\equiv\omega+\Omega$ is
the longitude of perihelion and $\Omega$ is the longitude of the
ascending node. The condition $j+j^{'}+l+l^{'}+m+m^{'}=0$ must be,
in general, fulfilled together with that which states that
$m+m^{'}$ must be an even number (positive, negative or zero).
Since now we are not looking at the secular terms, the
supplementary condition $l+l^{'}+m+m^{'}\neq 0$ must be fulfilled
as well.
It turns out that mixed terms proportional to, say, $ee^{'2}$
contribute to \rfr{short}. For Jupiter, the factor $2GM^{'}/n^2 a
a^{'}$ amounts nominally to $8.2258248\times 10^6$ m. The
uncertainty due to $\sigma_{Gm^{\rm Jup}}=2\times 10^9$ m$^3$
s$^{-2}$ is of the order of 10$^{-1}$ m, while $\sigma_{a^{\rm
Jup}}=675.1$ m yields an error of the order of $10^{-3}$ m.
It seems, then, reasonable to conclude that the indirect
perturbations on the semimajor axis should not pose severe
limitations to the proposed measurement.
\end{itemize}
\section{The (im)possibility of a
measurement with Earth artificial satellites and the Moon} Let us
now see if the gravitoelectric correction $P^{(\rm GE)}$ could be
measured by analyzing the motion of  Earth artificial and natural
satellites whose eccentricities are rather small.

From an observational point of view, by repeating the reasonings
of Section \ref{sensi} with \rfr{daa} and by considering that the
accuracy in the orientation of the axes of the International
Terrestrial Reference System \footnote{Its realization, the
International Terrestrial Reference Frame (ITRF) \citep{mcc04}, is
obtained by estimates of the coordinates and velocities of a set
of observing stations on the Earth.} (ITRS) is of the order
\footnote{See on the Internet
http://hpiers.obspm.fr/iers/bul/bulb/explanatory.html} of 3 mas,
it would be possible to measure $P^{(\rm GE)}$ over a sufficiently
high number of orbital revolutions. Indeed, for a typical
satellite orbit with $a=7\times 10^6$ m the post-Newtonian shift
over $2\pi$ would amount to 1 mas.

However, by assuming, e.g., $a=1.2270\times 10^7$ m, as for LAGEOS
whose rms post-fit residuals amount to $\sigma_a\sim 10^{-2}$ m or
better, the systematic error on the Keplerian period $P^{(0)}$
amounts to 1$\times 10^{-5} $ s, while $P^{(\rm GE)}$ is of the
order of $7\times 10^{-6}$ s only. It is important to note that
this uncertainty, opposite to the observational one, could not be
reduced by observing a large number of transits: indeed, the 1-cm
level of accuracy in estimating the semimajor axis comes just from
a data reduction over a time span of many orbits.

Let us, now, consider the possibility of using the very accurate
present and future data from the Lunar Laser Ranging (LLR)
technique for the Moon \citep{wil04}. Recalling that $a^{\rm
Moon}=3.84400\times 10^{8}$ m and $e^{\rm Moon}=0.0554$ (Standish
2001), the approximation of circular orbit would be adequate and
$P^{(\rm GE)}=4\times 10^{-5}$ s almost independently of $f_0$;
the gravitoelectric advance over 2$\pi$ amounts to 7.5 mas. By
assuming the present day centimeter accuracy in knowing the lunar
orbit, the systematic error related to the Keplerian period, which
cannot be reduced by accumulating a large number of revolutions,
would amount to $1\times 10^{-4}$ s. If the millimeter accuracy
level will be reached \citep{wil04}, the systematic error will be
reduced down to $1\times 10^{-5}$ s, i.e. 25$\%$ of the
gravitoelectric correction.

The major limiting factor both for the Moon and the artificial
satellites is the impact of the uncertainty in the Earth's $GM$
which amounts to $\sigma_{GM_{\oplus}}=8\times 10^{5}$ m$^3$
s$^{-2}$ \citep{mcc04}. Indeed, the ratio of the error in the
Keplerian period due to $\sigma_{GM_{\oplus}}$ to the
gravitoelectric term amounts to \eqi
\rp{\left.\sigma_{P^{(0)}}\right|_{\rm GM_{\oplus}}}{P^{(\rm
GE)}}=\rp{c^2(\sigma_{GM_{\oplus}})}{3(GM_{\oplus})^2}\
a=(1.5084\times 10^{-7} \ {\rm m^{-1}})\ a. \eqf Then, for a
typical satellite orbit with $a=7\times 10^6$ m this systematic
bias would be close to 100$\%$. For the Moon it would be 5800$\%$.
\section{The (im)possibility of a
measurement with exoplanets} A potentially interesting category of
celestial objects which fit the conditions of validity of
\rfr{perge} is represented by many exoplanets \footnote{See on the
WEB http://cfa-www.harvard.edu/planets/catalog.html.} recently
discovered with the transit method. Indeed, they are giant planets
following circular trajectories with orbital periods of the order
of 1 day, i.e. very close to their star. On the other hand, the
accuracy of the measurement of their periods with photometric
techniques is continuously increasing. Let us look at the very
recently analyzed OGLE-TR-132b \citep{bou04, mou04}. The relevant
orbital parameters are $M/M_{\odot}=1.35\pm 0.06$ (adopted for the
star), $a/{\rm A.U. }=0.0306\pm 0.0008$, $m^{\rm (c)}/m^{\rm
(Jup)}=1.19\pm 0.13$ and $P^{(\rm obs)}=1.46003645\times 10^5\
{\rm s}\pm 0.518$ s (measured for the planet over many orbital
revolutions). Unfortunately, $P^{(\rm GE)}=0.095$ s only.

The situation is even worse in the case of the pulsars' planetary
systems. For, e.g.,the 0.020 $m_{\oplus}$ planet orbiting
\footnote{See http://www.obspm.fr/encycl/1257+12.html} PSR 1257+12
at 0.19 A.U. the uncertainty in the observed orbital period is
259.2 s while $P^{(\rm GE)}$ is 0.24 s.

%

\section{Conclusions}
In this paper we have conducted a preliminary sensitivity analysis
about the possibility of measuring the post-Newtonian
gravitoelectric correction to the orbital period of a test body in
a Solar System scenario. The conclusions are summarized in Table
1.
\begin{table*}
 \centering
 \begin{minipage}{140mm}
  \caption{Observational and systematic errors affecting the measurement of
  the post-Newtonian gravitoelectric correction $P^{(\rm GE)}$ to the orbital period of Mercury, the Moon, the Earth artificial satellite
  LAGEOS, an exoplanet of the system OGLE-TR-132b and a planet of the PSR 1217+12 pulsar. All figures are in s. The circular
  orbit approximation has been adopted. The observational accuracy is to be intended over one orbital
  revolution. It is retrieved from the ratio of the accuracy, in
  mas, with which the axes of the ICRF and ITRS are known
  to the gravitoelectric shift over one orbital revolution.
Contrary to the systematic errors, the observational errors can be
reduced by observing a sufficiently large number of orbital
revolutions. Note that the observational uncertainty for the
exoplanet refers to many orbital revolutions. The systematic
errors are due to
   the Keplerian period $P^{(0)}$,  the indirect effect
  $\Delta P^{(0)}$ on it due to
  the perturbations in the semimajor axis,  the perturbation induced by the quadrupolar mass
  moment of the central source $P^{(J_2)}$,  the direct perturbation induced by the secular N-body interaction
  $P^{(\rm planets)}$. In regard to the indirect effects on the Keplerian period, $\sigma_{\Delta P^{(0)}}|_{\rm GE}$ and
  $\sigma_{\Delta P^{(0)}}|_{\rm planets}$ are induced by the short-period post-Newtonian and Newtonian effects on the semimajor axis. }
  \begin{tabular}{@{}llllll@{}}
  \hline
 & Mercury & Moon & LAGEOS & OGLE-TR-132b & PSR 1217+12\\
  \hline
$P^{(\rm GE)}$ & $2.9\times 10^{-1}$ & $4\times 10^{-5}$ &
$7\times 10^{-6}$ & $9.5\times 10^{-2}$ &$2.4\times 10^{-1}$\\
\hline $\sigma_{\rm obs}$ & $1\times 10^{-4}$ & $1.6\times
10^{-5}$ & $2\times 10^{-5}$ & $5.18\times 10^{-1}$&
$2.592\times 10^2$\\
\hline $\sigma_{P^{(0)}}|_a$ & $3\times 10^{-5}$ & $1\times
10^{-4}$ & $1\times 10^{-5}$ & \\
$\sigma_{P^{(0)}}|_{GM}$ & $2\times 10^{-4}$ & $2\times 10^{-3}$ &
$1\times 10^{-5}$ & \\
$\sigma_{P^{(J_{2})}}$ & $1.2\times 10^{-4}$ & & &\\
$\sigma_{P^{(\rm planets)}}$ & $\leq 10^{-7}$ & & &\\
$\sigma_{\Delta P^{(0)}}|_{\rm GE}$ & $\leq 10^{-11}$ & & &\\
$\sigma_{\Delta P^{(0)}}|_{\rm planets}$ & $\leq 10^{-7}$ & & &\\
\hline
\end{tabular}
\end{minipage}
\end{table*}

The post--Newtonian shift of Mercury's right ascension would
amount to 50 mas over one orbital revolution. Its orbit has been
assumed circular just in order to perform a preliminary
sensitivity analysis. In fact, the obtainable observational
accuracy  does not allow to neglect the impact of the factor of
\rfr{fattore} which depends on the eccentricity and the true
anomaly at the initial epoch.

Such  an effect should be measurable, in principle, for Mercury
with a $10^{-4}$  accuracy. This level is mainly set by the
systematic errors induced by the mismodelling in the Keplerian
period which should be subtracted.  Instead, the precision with
which the axes of the International Celestial Reference Frame are
known (0.02 mas) can be assumed as representative of the
observational accuracy. It yields a $10^{-4}$  accuracy over one
orbital revolution only. Of course, it could be increased by
observing as many transits as possible. Future improvements in
astrometry and in the knowledge of Mercury's orbital parameters
from the Messenger and BepiColombo mission should allow to further
increase the precision of this test by reducing the systematic
errors. BepiColombo is also expected to measure the Solar
quadrupole mass moment $J_2$ with an accuracy of few parts in
$10^{-9}$.


The present-day level of accuracy in knowing the orbits of the
best accurately tracked Earth artificial satellites rules out the
possibility of measuring such post-Newtonian effect in the
terrestrial gravitational field. For the Moon, a millimeter
accuracy in knowing its orbit, a goal which could be reached in
the next future, would yield a 25$\%$ systematic bias due to this
source of error. Instead, the present-day centimeter level in LLR
would not allow to sufficiently reduce the systematic error due to
the uncertainty in the Keplerian period. However, the uncertainty
in the present-day knowledge of the geocentric gravitational
constant $GM_{\oplus}$ would induce a systematic error of more
than 100$\%$ for every terrestrial artificial or natural
satellites.

The exoplanet scenario is not known with a sufficiently accuracy
from the point of view of the observational errors
%



\section*{Acknowledgments}
I am grateful to E. M. Standish (JPL) for his helpful
clarifications on the accuracy of ICRF and to S. Turyshev (JPL)
for his generous and strong encouragements and efforts to improve
this work. I also thank the anonymous referee for her/his careful
reading of the manuscript and her/his detailed comments.

\bsp

\label{lastpage}

\end{document}